\documentclass{article}

\usepackage{latexsym,amssymb,nath}

\newtheorem{proposition}{Proposition}

\newtheorem{definition}[proposition]{Definition}
\newtheorem{remark}[proposition]{Remark}
\newtheorem{example}[proposition]{Example}

\newtheorem{corollary}[proposition]{Corollary}
\newtheorem{construction}[proposition]{Construction}

\newenvironment{pf}{\removelastskip\par\medskip\noindent{\bf Proof }}%
 {\par\medskip}

\def\RR{\mathbb R}
\def\GL{\mathrm{GL}}

\def\SO{\mathrm{SO}}
\def\SU{\mathrm{SU}}
\def\S{\mathrm{S}}
\def\U{\mathrm{U}}
\def\e{\mathrm{e}}

\def\diag{\mathrm{diag}}
\def\pr{\mathrm{pr}}

\def\sy#1{\expandafter\csname #1\endcsname}
\def\newsy#1{\expandafter\def\expandafter\csname #1\endcsname}

\begin{document}

\title{Reducibility of zero curvature representations with
application to recursion operators}

\author{Michal Marvan\thanks{Mathematical Institute, Silesian University in 
  Opava, Bezru\v{c}ovo n\'am.~13, 746~01 Opava, Czech Republic.
 {\it E-mail}: Michal.Marvan@math.slu.cz}}

\date{}
\maketitle

\begin{abstract} 
We present a criterion of reducibility of a zero curvature representation
to a solvable subalgebra, hence to a chain of conservation laws.
Namely, we show that reducibility is equivalent to the existence of a section 
of the generalized Riccati covering.
Results are applied to conversion between Guthrie's and Olver's form of
recursion operators.
\end{abstract}

\section{Introduction}

To establish integrability of a nonlinear partial differential equation 
in the sense of soliton theory~\cite{A-S,T-F}, at least in two dimensions, 
one usually looks for a zero curvature representation (ZCR)~\cite{Z-S}, 
possibly in the form of a Lax pair~\cite{L}.
If depending on a non-removable (spectral) parameter, a ZCR may serve as 
a starting point of methods to derive infinitely many independent 
conservation laws and large classes of exact solutions.

However, certain ZCR's do not imply integrability because of specific
degeneracy, which does not even rule out possible dependence on one 
or more nonremovable parameters.
E.g., Calogero and Nucci~\cite{C-N} gave a formula to assign a
Lax pair to any nonlinear system possessing a single conservation law,
arguing that such systems are too abundant to be all integrable.
Recently Sakovich~\cite{Sak} observed that the Calogero--Nucci examples 
can be singled out by properties of their associated cyclic bases.
In particular, the `bad' ZCR's fail to generate an integrable hierarchy.

In this paper we postulate that a ZCR is degenerate if it takes values in 
a solvable Lie algebra or is gauge equivalent to such.
Even though some researchers are inclined to admit the relevance of such 
ZCR's to integrability, our results below seem to support the opposite 
opinion.

ZCR's taking values in an abelian algebra are well known to be equivalent 
to a set of local conservation laws (see~\cite[Sect. 3.2.c]{A-S}).
Using the Lie theorem on finite-dimensional representations of solvable 
algebras, we show in Sect.~4 rather easily that every ZCR that takes values 
in a solvable algebra is equivalent to a `chain' of nonlocal conservation 
laws.
This simple result renders, e.g., attempts to generate infinitely many 
independent conservation laws out of a degenerate ZCR rather unrealistic.

In Sect.~\ref{reducibility} we address the problem of detecting 
reducibility of a ZCR to a subalgebra, in particular, to a solvable one.
Purely algebraic criteria are insufficient since the Lie algebra a ZCR 
takes values in may be altered by gauge transformation.
On the other hand, when trying to find the reducing gauge matrix directly, 
we face a rather large underdetermined differential system.
Our idea is to employ an appropriate matrix decomposition, namely, 
the Gram or Gauss decomposition.
Earlier these decompositions were applied by Dodd and Paul~\cite{D,D-P} in 
the context of B\"acklund transformations.
A remarkable connection between decompositions and integrable systems emerged 
in numerical analysis \cite{Ch-N,D-N-T,W}. 

The last section is devoted to recursion operators, direct and inverse, for 
symmetries of integrable systems~\cite{B-K,K-K,Ol,V}.
In Olver's~\cite{O,Ol} formalism, a recursion operator is a linear
integro-differential operator $\Psi$, which maps symmetries to symmetries.
The standard way of inverting $\Psi$ consists in finding differential operators 
$K,L$ such that $\Psi = L \circ K^{-1}$;
then $\Psi^{-1} = K \circ L^{-1}$.
However, one encounters the problem of writing the inverse $L^{-1}$ as an 
integro-differential operator.
In the scalar case, $L$ may be put in the form 
$L = q_n D q_{n-1} D \cdots q_1 D q_0$, where the coefficients $q_i$ are 
expressible as quotients of wronskians of independent solutions $v_i$ of 
$L(v) = 0$
(see~\cite{Ri} for a simple derivation of this classical result,
equivalent to decomposability into first-order factors;
see \cite{Zh} for the matrix case). 
In our context, $q_i$ are nonlocal functions and finding them is considered
to be the most difficult part of the whole procedure.
Once $q_i$ are found, one can invert $L$ simply as 
$L^{-1}
 = q_0^{-1} D^{-1} q_1^{-1} \cdots D^{-1} q_{n-1}^{-1} D^{-1} q_n^{-1}$.
This is essentially the general scheme behind the 
works~\cite{G-H,L1,L2,L3,Q-S,R-L}.

Guthrie's~\cite{G} recursion operators resemble B\"acklund autotransformations 
for the linearized system and indeed can be interpreted this way
(see~\cite{M2});
their inversion is quite straightforward and does not require the
introduction of new nonlocalities. 
Moreover, Guthrie's operators do not suffer from the known abnormities, 
related to the fact that $D^{-1} \circ D = \hbox{id}$ fails to 
hold~(\cite{G-K-T,S-W}).
Let us also remind the reader that computing the `inverse' Guthrie operator 
starting from a known ZCR may turn out to be easier than computing the 
`direct' one (see~\cite{M-S}).
The conversion from Olver's to Guthrie's form was explained by 
Guthrie~\cite{G} himself, the result being further strengthened by 
Sergyeyev~\cite{Se}.
Concerning the backward conversion, the $x$-part of a Guthrie operator can 
be written as an integro-differential operator if the ZCR underlying it is 
lower triangular. 
A non-parametric ZCR can be made lower triangular at the cost of the 
introduction of appropriate nonlocalities. 
To introduce only few (respected) nonlocalities, 
we take into account a particular observation (already exploited in~\cite{M-S}) 
about the structure of Guthrie's recursion operators of integrable systems.

\section{Preliminaries}

Let $E$ be a system of nonlinear partial differential equations (PDE)
\begin{equation}
\label{eq}
F^l = 0
\end{equation}
on a number of functions $u^k$ in two independent variables $x,y$. 
Here each $F^l$ is a smooth function depending on a finite number of 
variables $x,y,u^k,u^k_x,u^k_y,\dots$, $u^k_I$,\dots, where $I$ 
stands for a symmetric multiindex over the two-element set of indices 
$\{x,y\}$.
Besides the {\it local\/} variables $x,y,u^k,u^k_I$, we shall also 
need non-local variables or pseudopotentials~\cite{W-E}, which may be 
introduced as additional variables satisfying a system of equations 
\begin{equation}
\label{z}
z^i_x = f^i, \quad z^i_y = g^i,
\end{equation}
where $f^i,g^i$ are functions depending on a finite number
of local variables as well as the pseudopotentials $z^j$;
we require that the system~(\ref{z}) be compatible as a consequence of 
(\ref{eq}).

Within their geometric theory of systems of PDE's, Krasil'shchik and 
Vinogradov~\cite{K-V} introduced the notion of a covering, which separates 
the invariant content of nonlocality from its coordinate presentation.
Pseudopotentials then correspond to a particular but arbitrary choice of 
coordinates along the fibres of the covering in question.
We recall the basic facts below; details we had to leave aside may be found 
in~\cite{K-V} and also in~\cite[Ch. 6]{B-V-V}.
Readers interested mainly in practical computations may skip the rest of
this section.

Let $J^\infty$ be an infinite jet space equipped with local jet 
coordinates $x,y$, $u^k,u^k_I$; the functions $F^l$ then may be interpreted 
as functions defined on $J^\infty$.
Since all our considerations are local, we simply let $J^\infty$ 
be the space of jets of sections of the trivial fibred manifold 
$Y \times M \to M$, where $M = \RR^2$, with $x,y$ being coordinates on 
$\RR^2$ and $u^k$ being coordinates on~$Y$.
On $J^\infty$, we have two distinguished commuting vector fields
$$
D_x = \frac{\partial}{\partial x}
 + \sum_{k,I} u^k_{Ix}\frac{\partial}{\partial u^k_I},
\qquad
D_y = \frac{\partial}{\partial y}
 + \sum_{k,I} u^k_{Iy}\frac{\partial}{\partial u^k_I},
$$ 
which are called {\it total derivatives}.

The {\it equation manifold} $\mathcal E$ associated with system~(\ref{eq}) 
is defined to be the submanifold in $J^\infty$ determined by the infinite
system of equations $F^l = 0$ and $D_I F^l = 0$ for $I$ running through 
all symmetric multiindices in $x,y$. 
The total derivatives $D_x,D_y$ are tangent to $\mathcal E$, therefore 
they admit a restriction to~$\mathcal E$.
In what follows, equations will be identified with equation manifolds 
equipped with the restricted total derivatives; this approach is indeed 
very practical and suitable for all needs to be encountered below.

Mappings between equation manifolds that commute with projections to the
base manifold $M$ and preserve the total derivatives will be called 
{\it morphisms} of equations; they map solutions to solutions (we shall not
use the general morphisms of {\it diffieties} which need not commute with the 
projections and only preserve the distributions generated by the total 
derivatives).
Bijective morphisms are called {\it isomorphisms}; their inverses are
isomorphisms, too.

A {\it covering} over an equation $\mathcal E$ consists of another equation 
$\mathcal E'$ and a surjective morphism $\mathcal E' \to \mathcal E$.

The system formed by Equation (\ref{eq}) and the $2k$ additional 
equations~(\ref{z}) generates a covering, where $\mathcal E'$ is the 
trivial vector bundle $\mathcal E \times \mathbb R^k$ and 
$z^1,\dots,z^k$ provide coordinates along $\mathbb R^k$.
In particular, the projection preserves the coordinates $x,y$.
If $f^i,g^i$ are functions defined on~$E'$ such that the vector fields 
\begin{equation}
\label{D'}
D'_x = D_x + \sum_{i=1}^k f^i \frac\partial{\partial z^i}, \qquad 
D'_y = D_y + \sum_{i=1}^k g^i \frac\partial{\partial z^i}
\end{equation}
commute (which is a geometric way of saying that Equations~(\ref{z}) are
compatible), then $\mathcal E'$ equipped with the vector fields (\ref{D'}) 
is a $k$-dimensional covering over $\mathcal E$. 
Recall from~\cite{K-V} that every finite-dimensional covering is locally of 
this form.

Two coverings $\mathcal E'$ and $\mathcal E''$ are said to be 
{\it isomorphic over $\mathcal E$} if there exists an isomorphism of the 
equations $\mathcal E' \cong \mathcal E''$ that commutes with the 
projections to~$\mathcal E$.
Isomorphic coverings result from invertible transformations of nonlocal 
variables.
A $k$-dimensional covering is said to be {\it trivial} if it is isomorphic 
to one with $f^i = g^i = 0$; such a covering is essentially a family of 
identical copies of $\mathcal E$. 

The simplest yet useful covering~(\ref{z}) may be associated with a single 
nontrivial conservation law $\alpha = f\,dx + g\,dy$, i.e., a pair of 
functions $f,g$ defined on $\mathcal E$ and satisfying $D_y f = D_x g$ on 
$\mathcal E$:

\begin{definition} \label{1_abelian_cov} \rm
A {\it one-dimensional abelian covering} associated with a conservation 
law~$\alpha = f\,dx + g\,dy$ is defined to be the trivial vector bundle
$\mathcal E \times \mathbb R \to \mathcal E$, equipped with total derivatives
$$
D'_x = D_x + f \frac\partial{\partial z}, \qquad 
D'_y = D_y + g \frac\partial{\partial z},
$$ 
where $z$ denotes the coordinate along $\mathbb R$.
\end{definition}

As $f,g$ do not depend on $z$, the vector fields $D'_x,D'_y$ on $\mathcal E'$ 
commute if and only if $D_y f = D_x g$.
The variable $z$ is called the {\it potential} of the conservation 
law~$\alpha$. 
We have $D'_x z = f$, $D'_y z = g$ or briefly $z_x = f$, $z_y = g$.

Recall that a conservation law is said to be {\it trivial\/} if there exists
a (local) function $h$ on $\mathcal E$ such that $f = D_x h$, $g = D_y h$.
A covering associated to a trivial conservation law is isomorphic to
a trivial covering through the invertible change of variables $z = z' + h$.

A covering $\bar{\mathcal E} \to \mathcal E$ is said to be 
{\it trivializing} for a conservation law $\alpha = f\,dx + g\,dy$, 
if the pullback $\bar\alpha$ of $\alpha$ along the projection 
$\bar{\mathcal E} \to \mathcal E$ is a trivial conservation law on 
$\bar{\mathcal E}$.
Obviously, the one-dimensional abelian covering associated with the 
conservation law $\alpha$ trivializes $\alpha$.

A general $n$-dimensional abelian covering is obtained by repeating the
construction of the one-dimensional abelian covering 
(cf. \cite[Sect.~IV]{W-E}):

\begin{definition} \label{abelian_cov} \rm
An $n$-dimensional covering $\tilde{\mathcal E}$ over $\mathcal E$ is said 
to be {\it abelian}, if

(1) either $n = 1$ and $\tilde{\mathcal E}$ is a one-dimensional abelian 
covering over $\mathcal E$ in the sense of Definition~\ref{1_abelian_cov};

(2) or $\tilde{\mathcal E}$ is a one-dimensional abelian covering over an 
$(n-1)$-dimensional abelian covering $\mathcal E'$ over $\mathcal E$.
\end{definition}

Let us remark that Khorkova~\cite{Kh} introduced the {\it universal abelian 
covering}, which need not be finite-dimensional.

\section{Zero-curvature representations}

Simplest pseudopotentials that are not potentials are associated with
non-degenerate zero-curvature representations. 
Let $\mathfrak g$ be a matrix Lie algebra (recall that according to the Ado 
theorem every finite-dimensional Lie algebra has a matrix representation).
By a {\it $\mathfrak g$-valued zero-curvature representation} (ZCR) for 
$\mathcal E$ we mean a $\mathfrak g$-valued one-form $\alpha = A\,dx + B\,dt$ 
defined on $\mathcal E$ such that 
\begin{equation} \label{mc}
D_y A - D_x B + [A,B] = 0
\end{equation}
holds on $\mathcal E$, which means that (\ref{mc}) holds as a consequence
of system~(\ref{eq}) (we do not insist that (\ref{mc}) necessarily reproduces 
system (\ref{eq}), which is normally required in integrability theory). 

Let $\mathcal G$ be the connected and simply connected matrix Lie group 
associated with~$\mathfrak g$.
Then for an arbitrary $\mathcal G$-valued function $S$, the form 
$\alpha^S = A^S\,dx + B^S\,dt$, where
\begin{equation}\label{gauge}
A^S = D_x S S^{-1} + S A S^{-1}, \qquad
B^S = D_y S S^{-1} + S B S^{-1}
\end{equation} 
is another ZCR, which is said to be {\it gauge equivalent} to the former.

A ZCR is said to be {\it trivial\/} if it is gauge equivalent to zero, i.e., 
if $A = D_x S S^{-1}$, $B = D_y S S^{-1}$.
A covering $\bar{\mathcal E} \to \mathcal E$ is said to {\it trivialize}
a ZCR $\alpha = A\,dx + B\,dy$ if the pullback $\bar\alpha$ of $\alpha$ 
along the projection $\bar{\mathcal E} \to \mathcal E$ is a trivial ZCR.

A trivializing covering for the ZCR $\alpha$ can be obtained in the 
following way.

\begin{proposition}
For every $\mathfrak g$-valued ZCR $\alpha$ on $\mathcal E$ there exists a 
covering $\pi_\alpha : \tilde{\mathcal E}_\alpha \to \mathcal E$ 
that trivializes $\alpha$. 
\end{proposition}

\begin{pf}
Let $\alpha = A\,dx + B\,dy$ be a ZCR, where $A$ and $B$ are $n \times n$ 
matrices belonging to the algebra $\mathfrak g$.
Put $\tilde{\mathcal E}_\alpha = \mathcal E \times G$, where $G$ is the 
matrix Lie group associated with~$\mathfrak g$.
Given an element $C \in \mathfrak g$, let us denote by $\xi_C$ the 
right-invariant vector field on $G$ corresponding to $C$.
Given a $\mathfrak g$-valued function $C$ on $\mathcal E$, let us denote by 
$\Xi_C$ the unique vector field on $\tilde{\mathcal E}_\alpha$ with the 
$\mathcal E$-component zero and the $G$-component equal to $\xi_C$, at each 
point of $\tilde{\mathcal E}_\alpha$.
Considering the vector fields
$$
\tilde D_x = D_x + \Xi_A, \qquad
\tilde D_y = D_y + \Xi_B
$$
on $\tilde{\mathcal E}_\alpha$, where $D_x, D_y$ are the total derivatives 
on~$\mathcal E$, let us show that $\smash{\tilde D_x, \tilde D_y}$ are the 
total derivatives for a trivializing covering 
$\pi_\alpha : \smash{\tilde{\mathcal E}_\alpha} \to \mathcal E$ 
of~$\alpha$.

Let $A = (a_{ij})$, $B = (b_{ij})$.
Let us first consider $G = \GL_n$ with its natural parametrization 
$\GL_n = \{(z_{ij}) \mid \det z_{ij} \ne 0\}$. 
We have 
$$
\Xi_A = \sum_{i,j,l} a_{ij} z_{jl} \frac{\partial}{\partial z_{il}}, \qquad
\Xi_B = \sum_{i,j,l} b_{ij} z_{jl} \frac{\partial}{\partial z_{il}}.
$$
Then $\tilde D_x,\tilde D_y$ commute since
\begin{eqnarray*}
[\tilde D_x,\tilde D_y]
& = & [D_x,D_y] + [D_x,\Xi_B] - [\Xi_A,D_y] + [\Xi_A,\Xi_B] \\
& = & \Xi_{D_x B - D_y A - [A,B]} \\
& = & 0.
\end{eqnarray*}
The same holds for arbitrary $G \subseteq \GL_n$, since the vector fields 
$\Xi_A,\Xi_B$ are tangent to $G$ whenever $A,B$ belong to~$\mathfrak g$.

Now denote by $\Theta$ the projection 
$\tilde{\mathcal E}_\alpha = \mathcal E \times G \to G$ 
viewed as a matrix-valued function on $\tilde{\mathcal E}_\alpha$.
Then $D_x \Theta = 0$ and therefore
$$
(\tilde D_x \Theta)_{\mu\nu}
 = (\Xi_A \Theta)_{\mu\nu}
 = \sum_{i,j,l} a_{ij} z_{jl} \frac{\partial}{\partial z_{il}} z_{\mu\nu}
 = \sum_j a_{\mu j} z_{j\nu} 
 = (A \Theta)_{\mu\nu}.
$$
Thus, $\tilde D_x \Theta \cdot \Theta^{-1} = A$
and similarly $\tilde D_y \Theta \cdot \Theta^{-1} = B$,
whence the pullback of $\alpha$ on \smash{$\tilde{\mathcal E}_\alpha$} 
is trivial.
\end{pf}

The system~(\ref{z}) corresponding to $\tilde{\mathcal E}_\alpha$ can be
compactly written in terms of a single matrix $\Theta$ as
\begin{equation}
\label{Theta}
\Theta_x = A \Theta, \qquad \Theta_y = B \Theta.
\end{equation}

%

Under the gauge transformation (\ref{gauge}), the matrix $\Theta$ becomes 
$S \Theta$. 
The coverings $\tilde{\mathcal E}_\alpha$ and $\tilde{\mathcal E}_{\alpha^S}$ 
are isomorphic via $\Theta \mapsto S \Theta$.

The trivializing covering $\pi_\alpha$ just constructed has the following
factorization property:

\begin{proposition} \label{factorization}
Let $p : \mathcal E' \to \mathcal E$ over $M$ be a trivializing covering 
for a ZCR $\alpha$ on $\mathcal E$.
Then there exists a morphism 
$p^\sharp : \mathcal E' \to \tilde{\mathcal E}_\alpha$
such that $\pi_\alpha \circ p^\sharp = p$.
\end{proposition}

\begin{pf}
Let $\alpha = A\,dx + B\,dy$.
Since $p$ is over $M$, we have $p^*\alpha = p^*A\,dx + p^*B\,dy$.
By assumption this is a trivial ZCR, whence $p^*A = D'_x S S^{-1}$ and 
$p^*B = D'_y S S^{-1}$ for a suitable $G$-valued function $S$ 
on~$\mathcal E'$.
Recall that fibres of the covering \smash{$\tilde{\mathcal E}_\alpha$} are
diffeomorphic to the Lie group $G$.
Therefore we can define a mapping
$p^\sharp : \mathcal E' \to \tilde{\mathcal E}_\alpha$
by the formula $\Theta \circ p^\sharp = S$, where, as above, $\Theta$ 
denotes the projection 
$\tilde{\mathcal E}_\alpha = \mathcal E \times G \to G$.
The mapping $p^\sharp$ is a morphism, since
$(\Theta \circ p^\sharp)_x = S_x = AS = A\Theta \circ p^\sharp$.
\end{pf}

\section{Lower triangular ZCR's} 
\label{lower triangular}

Let $\mathfrak t_n$ denote the algebra of matrices
\begin{equation}\label{triang}
(\begin{array}{ccccc}
a_{11} & 0 & \cdot & \cdot & 0 \\
a_{21} & a_{22} & 0 & \cdot & \cdot \\
a_{31} & a_{32} & a_{33} & \cdot & \cdot \\
\cdot & \cdot & \cdot & \cdot & 0 \\
a_{n1} & a_{n2} & a_{n3} & \cdot & a_{nn}
\end{array}).
\end{equation}
Denote by $\mathfrak t^{(k)}_n$, $k \ge 1$, the derived algebra formed by 
matrices satisfying $a_{ij} = 0$ whenever $i - j \lt k$. 

ZCR's with values in $\mathfrak t_n$ are, in a sense, equivalent to an
abelian covering. 

\begin{proposition} \label{p:triv_abel}
Every $\mathfrak t_n$-valued ZCR can be trivialized by means of an abelian 
covering of dimension $\le \frac12 n(n + 1)$.
\end{proposition}

\begin{pf}
Let $\alpha = A\,dx + B\,dy$ be a ZCR such that matrices $A$ and $B$ are lower 
triangular.
We shall construct an abelian covering $\mathcal E^{(n - 1)}$ in $n$ steps.

It follows from Equation (\ref{mc}) that
$\gamma_1 = a_{11}\,dx + b_{11}\,dy$, $\gamma_2 = a_{22}\,dx + b_{22}\,dy$, 
\dots, 
$\gamma_n = a_{nn}\,dx + b_{nn}\,dy$ are conservation laws.
Let us denote by $\mathcal E^{(0)}$ the associated abelian covering 
with potentials $h_1, \dots, h_n$ satisfying
$$
h_{i,x} = a_{ii}, \quad
h_{i,y} = b_{ii} \qquad \text{for $i = 1,\dots,n$}.
$$
Then
$$
H = (\begin{array}{ccccc}
\e^{-h_1} & 0 & 0 & \cdot & 0 \\
0 & \e^{-h_2} & 0 & \cdot & 0 \\
0 & 0 & \e^{-h_3} & \cdot & 0 \\
\cdot & \cdot & \cdot & \cdot & \cdot \\
0 & 0 & 0 & \cdot & \e^{-h_n}
\end{array}),
$$
is a matrix defined on $\mathcal E^{(0)}$, with the property that
all diagonal entries of the gauge equivalent matrix $A' = A^H$ vanish:
\begin{equation}
\label{triangi}
A' = (\begin{array}{cc@{\kern 3ex}c@{\kern 1.2ex}c@{\kern 1.2ex}c}
0 & \cdot & \cdot & 0 & 0 \\
a'_{21} & 0 & \cdot & \cdot & 0 \\
a'_{31} & a'_{32} & 0 & \cdot & \cdot \\
\cdot & \cdot & \cdot  & \cdot & \cdot \\
a'_{n1} & a'_{n2} & \cdot & a'_{n,n-1} & 0
\end{array}),
\end{equation}
and similarly for $B'$.
Hence, $A',B'$ take values in $t^{(1)}_n$.

By the same Equation~(\ref{mc}), $\gamma'_2 = a'_{21}\,dx + b'_{21}\,dy$, 
$\gamma'_3 = a'_{32}\,dx + b'_{32}\,dy$, \dots, 
$\gamma'_n = a'_{n-1,n}\,dx + b'_{n-1,n}\,dy$ 
are conservation laws on $\mathcal E^{(0)}$.
Let us introduce a covering $\mathcal E'$ over $\mathcal E^{(0)}$ 
with potentials $h'_2, \dots, h'_n$ satisfying 
$$
h'_{i,x} = a'_{i,i-1}, \quad
h'_{i,y} = b'_{i,i-1} \quad \text{for $i = 2,\dots,n$}.
$$
Denoting
$$
H' = (\begin{array}{c@{\kern 1.5ex}cccc}
1 & 0 & \cdot & 0 & 0 \\
-h'_2 & 1 & \cdot & \cdot & 0 \\
0 & -h'_3 & 1 & \cdot & \cdot \\
\cdot & \cdot & \cdot & \cdot & \cdot \\
0 & \cdot & 0 & -h'_n & 1
\end{array}),
$$
we see that the gauge equivalent matrices $A'' = A'{}^{H'}$ and 
$B'' = B'{}^{H'}$ take values in $t^{(2)}_n$ now.
Compared with (\ref{triangi}), $A''$ and $B''$ have one more subdiagonal of 
zeroes.
The next step is similar:
$\gamma''_3 = a''_{31}\,dx + b''_{31}\,dy$, 
$\gamma''_4 = a''_{42}\,dx + b''_{42}\,dy$, \dots, 
$\gamma''_n = a''_{n-2,n}\,dx + b''_{n-2,n}\,dy$ are conservation laws on 
$\mathcal E'$.
Let us introduce a covering $\mathcal E''$ over $\mathcal E'$ with 
potentials $h''_2, \dots, h''_n$ satisfying 
$$
h''_{i,x} = a''_{i,i-2}, \quad
h''_{i,y} = b''_{i,i-2} \quad \text{for $i = 3,\dots,n$}.
$$
Denoting
$$
H'' = (\begin{array}{c@{\kern 1ex}cc@{\kern 1.7ex}c@{\kern 2.8ex}
 c@{\kern 3.5ex}c}
1 & 0 & \cdot & 0 & 0 & 0 \\
0 & 1 & \cdot & \cdot & 0 & 0 \\
-h''_3 & 0 & 1 & \cdot & \cdot & 0 \\
0 & -h''_4 & 0 & 1 & \cdot & \cdot \\
\cdot & \cdot & \cdot & \cdot & \cdot & \cdot \\
0 & \cdot & 0 & -h''_n & 0 & 1
\end{array})
$$
we observe that $A''' = A^{\prime\prime H''}$, $B''' = B^{\prime\prime H''}$
take values in $t^{(3)}_n$, and so on.
Continuing the process until $A^{(n)}$, $B^{(n)}$ become zero, we end up 
with a sequence of $\frac 12 n(n + 1)$ conservation laws
\begin{equation}
\label{e:chain}
\begin{array}{lllcl}
\gamma_1&\gamma_2&\gamma_3&\dots&\gamma_n \\ 
&\gamma'_2&\gamma'_3&\dots&\gamma'_n \\
&&\gamma''_3&\dots&\gamma''_n \\
&&&\dots \\ 
&&&\gamma^{(n-2)}_{n-1}&\gamma^{(n-2)}_n \\
&&&&\gamma^{(n-1)}_n,
\end{array}
\end{equation}
where
(a) $\gamma_1,\dots,\gamma_n$ are conservation laws on $\mathcal E$;
(b) $\gamma^{(n-\iota)}_{n-\iota+1}, \dots, \gamma^{(n-\iota)}_n$ 
are conservation laws defined on the abelian covering 
$\mathcal E^{(n-\iota-1)}$ 
associated with the conservation laws of all the previous levels.

Finally, $\alpha^{H H' \cdots H^{(n-1)}} = \alpha^{(n)} = 0$, where 
each $H^{(\iota)}$ is defined on~$\mathcal E^{(\iota)}$.
Summing up, the covering $\mathcal E^{(n - 1)}$ trivializes~$\alpha$.
\end{pf}

The sequence (\ref{e:chain}) will be called an {\it $n$-fold chain of 
conservation laws}.

\begin{proposition}
Let $\alpha$ be a $\mathfrak t_n$-valued ZCR, then the associated covering
$\pi_\alpha$ is isomorphic to an abelian covering of dimension 
$\le \frac12 n(n + 1)$.
\end{proposition}

\begin{pf}
According to Proposition~\ref{p:triv_abel}, there is an abelian covering 
$p : \mathcal E^{(n - 1)} \to \mathcal E$ that is trivializing for 
$\alpha$; 
namely, we have $\alpha^K = 0$, where $K = H H' \cdots H^{(n-1)}$
(see proof of Proposition~\ref{p:triv_abel}).
Hence, \smash{$\alpha = 0^{K^{-1}}$} and, according to 
Proposition~\ref{factorization}, there is a morphism 
$p^\sharp : \mathcal E^{(n - 1)} \to \tilde{\mathcal E}_\alpha$, given 
by $\Theta = K^{-1}$.
Here $\Theta$ represents the totality of coordinates along the fibres of 
the covering $\tilde{\mathcal E}_\alpha$, while $K$ is parametrised by
coordinates $h_s^{(\iota)}$ along the fibres of the covering 
$\mathcal E^{(n - 1)}$.
It follows that $p^\sharp$ is bijective on the fibres, hence isomorphism.
\end{pf}

\section{Reducibility}
\label{reducibility}

A $\mathfrak g$-valued ZCR is said to be {\it reducible} if it is gauge 
equivalent to a ZCR taking values in a proper subalgebra 
$\mathfrak h \subset \mathfrak g$;
otherwise it is said to be {\it irreducible}.

Let $\mathfrak h \subset \mathfrak g$ be a subalgebra.
We present a simple criterion for reducibility of a $\mathfrak g$-valued ZCR
to $\mathfrak h$.
Let $\mathcal H \subset \mathcal G$ be the Lie subgroup 
corresponding to the subalgebra $\mathfrak h$.
We call $\mathcal H$ a {\it right factor} if there exists a submanifold 
$\mathcal K \subset \mathcal G$ (possibly with singularities) such that the 
multiplication map
\begin{equation}
\label{kh}
\mu : \mathcal K \times \mathcal H \to \mathcal G, \qquad (K,H) \mapsto K H
\end{equation}
is a surjective local diffeomorphism.
The manifold $\mathcal K$ will be called a {\it cofactor}.
By surjectivity, every element $S \in \mathcal G$ can be decomposed as a 
product $S = K H$, where $K \in \mathcal K$ and $H \in \mathcal H$, 
possibly non-uniquely.
The map $\mu$ being a local diffeomorphism, $\mathcal K$ has the minimal 
possible dimension $\dim \mathcal K = \dim \mathcal G - \dim \mathcal H$.
If $\mathcal H$ is closed, then the assignment $K \mapsto K \mathcal H$ 
defines a local diffeomorphism of $\mathcal K$ onto the homogeneous space 
$\mathcal G/\mathcal H$.

\begin{proposition} \label{red}
Under the above notation, a $\mathfrak g$-valued ZCR $\alpha$ on $\mathcal E$ 
is reducible to the subalgebra $\mathfrak h$ if and only if there exists 
a local $\mathcal K$-valued matrix function $K$ on $\mathcal E$ such that 
$\alpha^K$ lies in $\mathfrak h$.
\end{proposition}

\begin{pf}
The gauge equivalence with respect to $H \in \mathcal H$ preserves the 
subalgebra $\mathfrak h$.
Therefore, the gauge-equivalent ZCR $\alpha^S = (\alpha^K)^H$ lies in 
$\mathfrak h$ if and only if $\alpha^K$ lies in $\mathfrak h$. 
\end{pf}

Otherwise said, if a ZCR is reducible to $\mathfrak h$, then the corresponding
gauge matrix can be found in $\mathcal K$.
Understandably, different choices of the cofactor $\mathcal K$ may lead to 
different reducibility criteria. 

In this paper we are primarily interested in detecting reducibility to a 
solvable subalgebra.
By the well-known Lie theorem (\cite[Sect.~9.2]{F-H}), every finite-dimensional 
representation of a solvable Lie algebra is equivalent to a representation 
by lower triangular matrices.
Hence, every ZCR reducible to a solvable subalgebra is reducible to 
$\mathfrak t_n$ (and can be trivialized using an abelian covering according to
Proposition~\ref{p:triv_abel}).

Let us therefore apply Proposition~\ref{red} to $\mathfrak h = \mathfrak t_n$.
There are two standard ways to make $\mathfrak t_n$ a right factor in 
$\mathfrak{gl}_n$.

\paragraph*{The QR or Gram decomposition} is an alternative formulation of 
the famous Gram--Schmidt ortogonalization algorithm. 
Namely, every $n \times n$ complex matrix $A$ can be decomposed as a 
product $A = QR$, where $Q \in \SU_n$ and $R \in \mathfrak t_n$
\cite[Ch. 6, Sect. 1.9]{Naj}. 
Proposition~\ref{red} then yields

\begin{proposition} \label{QRred}
A real {\rm(}complex\/{\rm)} ZCR $\alpha$ on $\mathcal E$ is reducible 
to lower triangular if and only if there exists an $\SO_n$-valued 
{\rm(}$\SU_n$-valued\/{\rm)}
function $K$ on $\mathcal E$ such that $\alpha^K$ is lower triangular.
\end{proposition}

However, the factors $Q$ and $R$ are unique up to a unimodular diagonal 
multiplier: $QR = Q\Theta \cdot \Theta^{-1}R$, where 
$\Theta = \diag(\theta_1,\dots,\theta_n) \in
 \S(\U_1 \times \cdots \times \U_1)$, i.e., $|\theta_\iota| = 1$ and
$\prod_{\iota=1}^n \theta_\iota = 1$.
Thus, the mapping (\ref{kh}) is not a local diffeomorphism unless it is
restricted to a suitable immersion of the orbit space 
$\SU_n/\S(\U_1 \times \cdots \times \U_1)$ into $\SU_n$.
In the real case we have $\theta_\iota = \pm 1$ and we get a $2^{n-1}$-to-one 
local diffeomorphism (\ref{kh}) with $\mathcal K = \SO_n$.

\paragraph*{The LU or Gauss decomposition} can be derived from the Gaussian 
elimination algorithm.
The following result is well known~(\cite[Ch. 6, Sect. 1.8]{Naj}):

\begin{proposition} \label{PLU}
For every non-singular matrix $A$ there exist matrices $P,U,L$ such that 
$A = PUL$, $L$ is lower triangular, $U$ is upper triangular with
diagonal entries equal to $1$, and $P$ is a permutation matrix.
The matrix $P$ can be omitted if and only if all principal minors of the 
matrix $A$ are nonzero.
\end{proposition}

(Recall that Gaussian elimination may require row swapping, which is where 
the permutation matrix $P$ comes from.)
Let $\mathcal K$ denote the set of all products $PU$ where $P$ is a 
permutation matrix and $U$ is an upper triangular matrix with diagonal 
entries equal to $1$.
Then $K$ is a union of $n!$ intersecting submanifolds, labelled by
permutation matrices $P$.
Compared to the QR-decomposition, each of the $n!$ submanifolds is easier 
to parametrize than $\SO$ or $\SU$.

\begin{proposition} \label{LUred}
A ZCR $\alpha$ on $\mathcal E$ is reducible to lower triangular if and only 
if there exists a permutation matrix $P$ and a matrix-valued function 
\begin{equation}
\label{H}
H = (\begin{array}{cccc@{\quad}c}
1 & h_{12} & h_{13} & \cdot & \cdot \\
0 & 1 & h_{23} & \cdot & \cdot \\
0 & 0 & 1 & \cdot & \cdot \\
\cdot & \cdot & \cdot & \cdot & \cdot \\
0 & \cdot & \cdot & 0 & 1
\end{array})
\end{equation} 
on $\mathcal E$ such that $\alpha^{PH}$ is lower triangular.
\end{proposition}

Before looking more closely at low values of $n$, we make a general
remark to the effect that every $\mathfrak{gl}_n$-valued ZCR is reducible to 
$\mathfrak{sl}_n$:

\begin{remark} \rm
\label{sl}
A $\mathfrak{gl}_n$-valued ZCR is decomposable into an 
$\mathfrak{sl}_n$-valued ZCR (traceless summand) and a conservation law
(the trace).
\end{remark}

\subsection{The case of $n = 2$}

When $n = 2$, the reducibility condition corresponding to the QR
decomposition is:

\begin{proposition} \label{QRgl2}
A $\mathfrak{gl}_2$-valued ZCR 
$$
\alpha = A\,dx + B\,dy
 = (\begin{array}{cr} a_{11} & a_{12} \\ a_{21} & a_{22}
   \end{array}) dx
 + (\begin{array}{cr} b_{11} & b_{12} \\ b_{21} & b_{22}
   \end{array}) dy
$$
is reducible to lower triangular 
if and only if there exists a function $\phi$ on $\mathcal E$ that is a 
solution of the system
\begin{equation}
  \label{eqQRgl2}
D_x \phi = -a_{12} \cos^2 \phi + (a_{11} - a_{22}) \sin \phi \cos \phi
 + a_{21} \sin^2 \phi, 
\\
D_y \phi = -b_{12} \cos^2 \phi + (b_{11} - b_{22}) \sin \phi \cos \phi
 + b_{21} \sin^2 \phi.
\end{equation}
\null
\end{proposition}

\begin{pf}
An arbitrary $\SO_2$ matrix is 
$$
K = (\begin{array}{rr} 
 \cos \phi & \sin \phi \\ 
 -\sin \phi & \cos \phi 
\end{array}).
$$
By Proposition \ref{QRred}, the ZCR $\alpha$ is reducible to lower 
triangular if and only if $\alpha^K$ is lower triangular, which is 
exactly the meaning of conditions~(\ref{eqQRgl2}).
\end{pf}

The reducibility conditions corresponding to the LU
decomposition are:

\begin{proposition} \label{ULgl2}
A $\mathfrak{gl}_2$-valued ZCR 
$$
\alpha = A\,dx + B\,dy
 = (\begin{array}{cr} a_{11} & a_{12} \\ a_{21} & a_{22}
   \end{array}) dx
 + (\begin{array}{cr} b_{11} & b_{12} \\ b_{21} & b_{22}
   \end{array}) dy
$$
on $\mathcal E$ is reducible to lower triangular 
if and only if 

1.
either there exists a local function $p$ on $\mathcal E$ such that
\begin{equation}
 \label{eqLUgl2}
D_x p = -a_{12} + (a_{11} - a_{22}) p + a_{21} p^2, 
\\
D_y p = -b_{12} + (b_{11} - b_{22}) p + b_{21} p^2;
\end{equation}

2.
or $A,B$ are upper triangular:
$$
a_{21} = b_{21} = 0.
$$
\end{proposition}

\begin{pf}
An arbitrary $\mathcal K$-valued function is $K = P U$, where
$$
U = (\begin{array}{rr} 
 1 & p \\ 
 0 & 1 
\end{array})
$$
and $P$ is one of the two $2 \times 2$ permutation matrices
$$
P_{12} = (\begin{array}{rr} 
 1 & 0 \\ 
 0 & 1 
\end{array}), \qquad
P_{21} = (\begin{array}{rr} 
 0 & 1 \\ 
 1 & 0 
\end{array}).
$$
Subcases 1 and 2 correspond to the choices $P = P_{12}$ and $P = P_{21}$,
respectively, and express the conditions that $A^{P U}$, $B^{P U}$ 
be lower triangular.
\end{pf}

Recall that a {\it quadratic} or {\it Riccati pseudopotential\/} $p$ 
associated to an $\mathfrak{gl}_2$-valued ZCR $\alpha$ is defined by the 
compatible system
\begin{equation}
\label{qupsp}
p_x = -a_{12} + (a_{11} - a_{22}) p + a_{21} p^2, 
\\
p_y = -b_{12} + (b_{11} - b_{22}) p + b_{21} p^2,
\end{equation}
which are essentially Equations (\ref{eqLUgl2}).
The system (\ref{qupsp}) being compatible, let us introduce the corresponding 
one-dimensional {\it Riccati covering}.
Proposition~\ref{ULgl2} then says that a non-upper-triangular ZCR is 
reducible to lower triangular if and only if the Riccati covering has a 
local section.

\begin{remark} \rm
Obviously, Equations (\ref{eqLUgl2}) and (\ref{eqQRgl2}) are not independent,
the explicit mapping of their solutions being $p = \tan\phi$ for 
$\phi \ne (2k + 1)\pi/2$.
Recently Reyes~\cite{Re} pointed out a geometric interpretation of the same
correspondence in terms of pseudospherical equations.
\end{remark}

\begin{example} \rm \label{Burgers}
The Burgers equation $u_t = u_{xx} + u u_x$ is well known to be integrable 
via the Cole--Hopf transformation, which relates its solutions with those
of the heat equation~\cite[Sect. 3.1]{A-S}.
Of the several Lax pairs that have been found all turn out to be degenerate.
Let us consider one example~\cite{Nu,We}, where the lower triangular 
representation could not be obtained by purely Lie algebraic methods:
\begin{eqnarray*}
\alpha &=& 
(\begin{array}{cc} 0 & 1 \\ 
-\frac{1}{4} u_x + \frac{1}{16} (u + \lambda)^2 & 0 
\end{array})
\,dx
\\ && + 
(\begin{array}{cc} 
-\frac{1}{4} u_x & \frac{1}{2} (u - \lambda) \\ 
-\frac{1}{4} u_{xx} - \frac{1}{8} (u - \lambda) u_x
 + \frac{1}{32} (u - \lambda) (u + \lambda)^2 & 
\frac{1}{4} u_x \end{array})
\,dt
\end{eqnarray*}
In this case, Equations (\ref{eqLUgl2}) have a local solution 
$p = 4/(u + \lambda)$, hence
$$
(\begin{array}{cc} 
1 & 4/(u + \lambda) \\ 0 & 1 
\end{array})
$$
is a gauge matrix to make the ZCR $\alpha$ lower triangular.
\end{example}

That the Burgers equation has no irreducible $\mathfrak{gl}_2$-valued ZCR 
follows from the recent classification of second-order evolution 
equations possessing an $\mathfrak{sl}_2$-valued ZCR~\cite{M3} and 
Remark~\ref{sl}.
The non-existence of an irreducible ZCR of Wahlquist--Estabrook 
type for arbitrary $n$ was proved Dodd and Fordy \cite{D-F} who
established solvability of the Wahlquist--Estabrook prolongation algebra 
of the Burgers (and also of the Kaup) equation.

\begin{example} \rm
The Calogero--Nucci example~\cite{C-N} of a ZCR that exists for every equation 
possessing a conservation law $f_t = g_x$:
\begin{equation}
\label{ }
(\begin{array}{c@{\quad}c} 
0 & 1 \\ 
\displaystyle \sy{eta} \frac{f_x}{f} + \sy{lambda} f^2
 + \sy{eta} \sy{mu} f - \sy{eta}^2
 & \displaystyle\frac{f_x}{f} + \sy{mu} f - 2 \sy{eta}
\end{array})\,dx
\\\qquad + 
(\begin{array}{c@{\quad}c} 
\displaystyle\sy{eta} \frac{g}{f} + \sy{nu} & \displaystyle\frac{g}{f} \\ 
\displaystyle\frac{\sy{eta} g_x}{f} + \sy{lambda} f g
 + \sy{eta} \sy{mu} g - \sy{eta}^2 \frac{g}{f}
 & \displaystyle\frac{g_x}{f} + \sy{mu}  g - \sy{eta} \frac{g}{f} + \sy{nu}
\end{array})\,dy
\end{equation}
where $\eta,\lambda,\mu,\nu$ are arbitrary constants.
This ZCR is reducible, which follows from Proposition~\ref{ULgl2}
along with explicit formulas for its reduction.
Indeed, we have Subcase 1 again and one easily finds a local solution
$$
p = \frac{1}{2} 
\frac{(\mu + \sqrt{\mu^2 + 4 \lambda}) f - 2 \eta}
{\lambda f^2 + \eta \mu  f - \eta^2}
$$
of Equations (\ref{eqLUgl2}).
Hence, the above ZCR is reducible to lower triangular.

Continuing the reduction further, one finally arrives at an abelian subalgebra.
Namely, if $p$ is as above and
\begin{eqnarray*}
q &=& \frac{\lambda f^2 + \eta \mu f - \eta^2}
         {\sqrt{\mu^2 + 4 \lambda}\,f},
\\
r &=& 
 \frac{(\lambda f^2 + \eta \mu f - \eta^2)
       \bigl(2 \lambda f
             + (\mu - \sqrt{\mu^2 + 4 \lambda})\eta
       \bigr)}
 {2 \lambda f + (\mu + \sqrt{\mu^2 + 4 \lambda}) \eta},
\end{eqnarray*}
then the product of gauge matrices 
$$
(\begin{array}{cc} \sqrt r/f & 0 \\
  0 & 1/\sqrt r \end{array})
(\begin{array}{cc} 1 & 0 \\ q & 1 \end{array})
(\begin{array}{cc} 1 & p \\ 0 & 1 \end{array})
$$
takes the ZCR to the diagonal form
$$
(\begin{array}{cc} 
\frac{1}{2}(\mu - \sqrt{\mu^2 + 4 \lambda})f - \eta & 
0 
\\ 
0 &
\frac{1}{2}(\mu + \sqrt{\mu^2 + 4 \lambda})f - \eta
\end{array})\,dx 
\\\qquad
 + (\begin{array}{cc} 
\frac{1}{2}(\mu - \sqrt{\mu^2 + 4 \lambda}) g + \nu
&0 \\ 0&
\frac{1}{2}(\mu + \sqrt{\mu^2 + 4 \lambda}) g + \nu
\end{array})\,dy,
$$
which is manifestly equivalent to the conservation law $f\,dx + g\,dt$.
\end{example}

\subsection{The case of $n \ge 3$}

For $n \ge 3$, the QR approach is impractical due to relative clumsiness of 
the parametrisation of $\SO_n$ by generalized Euler angles.
On the other hand, the LU criteria come out subdivided into as much as $n!$ 
subcases, one for each of the $n!$ permutation matrices~$P$.

For every $n$, the case of general position occurs when all principal 
minors of the gauge matrix $K$ are nonzero.
Then the permutation matrix $P$ equals the identity matrix and we can 
derive explicit formulas that generalize (\ref{eqLUgl2}) to arbitrary~$n$.

\begin{proposition} \label{ULgln}
A $\mathfrak{gl}_n$-valued ZCR 
$\alpha = A\,dx + B\,dy$, where $A = (a_{ij})$ and $B = (b_{ij})$, 
is reducible to lower triangular by means of a gauge matrix with nonzero
principal minors if and only if the system 
\begin{equation}
\label{eqLUsln}
\begin{array}{ll} 
D_x p_{kl} 
 = & \displaystyle -\sum_{
         0 \le r \le n - 1 \\
         i_0 \lt i_1 \lt \cdots \lt i_r = l
         }
(-1)^r a_{k i_0} p_{i_0 i_1} p_{i_1 i_2} \dots  p_{i_{r-1} i_r}
\\ 
   & \displaystyle - \sum_{
         0 \le r \le n - 1 \\
         k \lt j \\
         i_0 \lt i_1 \lt \cdots \lt i_r = l
         }
(-1)^r p_{kj} a_{ji_0} p_{i_0 i_1} p_{i_1 i_2} \dots  p_{i_{r-1} i_r},
\\ 
D_y p_{kl} 
 = & \displaystyle -\sum_{
         0 \le r \le n - 1 \\
         i_0 \lt i_1 \lt \cdots \lt i_r = l
         }
(-1)^r b_{k i_0} p_{i_0 i_1} p_{i_1 i_2} \dots  p_{i_{r-1} i_r}
\\ 
   & \displaystyle - \sum_{
         0 \le r \le n - 1 \\
         k \lt j \\
         i_0 \lt i_1 \lt \cdots \lt i_r = l
         }
(-1)^r p_{kj} b_{ji_0} p_{i_0 i_1} p_{i_1 i_2} \dots  p_{i_{r-1} i_r}
\end{array}
\end{equation}
on $\frac 12 (n - 1) n$ unknown functions $p_{kl}$, $k \lt l$, has a 
local solution.
\end{proposition}
\null

\begin{pf}
According to Proposition~\ref{PLU}, every gauge matrix $S$ with nonzero 
principal minors decomposes as $S = LU$, with $L$ lower triangular~and
$$
U = (\begin{array}{ccccc}
1 & p_{12} & p_{13} & \dots & p_{1n} \\
0 & 1 & p_{23} & \dots & p_{2n} \\
0 & 0 & 1 & \dots & p_{3n} \\
\vdots & \vdots & \vdots & & \vdots \\
0 & 0 & 0 & \dots & 1 
\end{array}).
$$
The inverse of $U$ is
$$
U^{-1} = (\begin{array}{ccccc}
1 & q_{12} & q_{13} & \dots & q_{1n} \\
0 & 1 & q_{23} & \dots & q_{2n} \\
0 & 0 & 1 & \dots & q_{3n} \\
\vdots & \vdots & \vdots & & \vdots \\
0 & 0 & 0 & \dots & 1 
\end{array}),
$$
where
\begin{eqnarray*}
q_{ij} &=& \sum_{
          1 \le r \le n - 1 \\
          i = i_0 \lt i_1 \lt \cdots \lt i_r = j
          }
(-1)^r p_{i_0 i_1} p_{i_1 i_2} \dots  p_{i_{r-1} i_r} \\
 &=& (-1)^{i+j} \left|\begin{array}{cccccc}
p_{i,i+1} & p_{i,i+2} & \cdot & \cdot & p_{i,j-1} & p_{i,j} \\
1 & p_{i+1,i+2} & p_{i+1,i+3} & \cdot & \cdot & p_{i+1,j} \\
0 & 1 & p_{i+2,i+3} & p_{i+2,i+4} & \cdot & \cdot \\
\cdot & \cdot & 1 & \cdot & \cdot & \cdot \\
0 & \cdot & \cdot & \cdot & p_{j-2,j-1} & p_{j-2,j} \\
0 & 0 & \cdot & \cdot & 1 & p_{j-1,j} 
\end{array}\right|,
\end{eqnarray*}
since $q_{kl} + \sum_{k \lt i \lt l} p_{ki} q_{il} + p_{kl} = 0$
whenever $k \lt l$.
Let us consider the gauge equivalent matrix $A^U = U_x U^{-1} + U A U^{-1}$.
Terms that contain total derivatives $D_x p_{ij}$ can occur only in the first
summand, which is
$$
U_x U^{-1} = (\begin{array}{ccccc}
0 & z_{12} & z_{13} & \dots & z_{1n} \\
0 & 0 & z_{23} & \dots & z_{2n} \\
0 & 0 & 0 & \dots & z_{3n} \\
\vdots & \vdots & \vdots & & \vdots \\
0 & 0 & 0 & \dots & 0 
\end{array}),
$$
where
\begin{eqnarray*}
z_{kl} 
 &=&\sum_{
          1 \le r \le n - 1 \\
          k = i_0 \lt i_1 \lt \cdots \lt i_r = l
          }
(-1)^{r-1} D_x p_{i_0 i_1} \cdot p_{i_1 i_2} \dots  p_{i_{r-1} i_r} 
\\
 &=& (-1)^{k+l+1} \left|\begin{array}{cccccc}
D_x p_{k,k+1} & D_x p_{k,k+2} & \cdot & D_x p_{k,l-1} & D_x p_{k,l} \\
1 & p_{k+1,k+2} & p_{k+1,k+3} & \cdot & p_{k+1,l} \\
0 & 1 & p_{k+2,k+3} & \cdot & \cdot \\
\cdot & \cdot & \cdot & \cdot & \cdot \\
0 & 0 & \cdot & 1 & p_{l-1,l} 
\end{array}\right|
\end{eqnarray*}
for all $k \lt l$.
Denoting $A^U =: A' = (a'_{ij})$, we have
$$
a'_{kl} := z_{kl} + a_{kl} + \sum_{j \lt l} a_{kj} q_{jl}
 + \sum_{k \lt i \\ j \lt l} p_{ki} a_{ij} q_{jl}
 + \sum_{k \lt i} p_{ki} a_{il}.
$$ 
The condition of $A'$ being lower triangular, $a'_{kl} = 0$ for all $k \lt l$, 
constitutes a system of equations in total derivatives $D_x p_{ij}$.
The equivalent system resolved with respect to the derivatives is 
$a'_{kl} + \sum_{k \lt h \lt l} a'_{kh} p_{hl} = 0$, since derivatives occur 
only in the summands containing $z_{ij}$, which are 
$z_{kl} + \sum_{k \lt h \lt l} z_{kh} p_{hl} = D_x p_{kl}$.
The remaining summands then simplify to the expressions given in the statement
of the proposition.
\end{pf}

\subsection{The generalized Riccati covering}

A tedious computation shows that Equations~(\ref{eqLUsln}) are compatible, 
meaning that there are no integrability conditions resulting from the 
equalities $D_{xy} p_{kl} = D_{yx} p_{kl}$.
This implies the existence of a covering associated with a ZCR which 
naturally generalizes the Riccati covering.
Similar result holds for more general types of decomposition, too.

Let a subalgebra $\mathfrak k \subseteq \mathfrak g$ be a direct complement 
to the subalgebra~$\mathfrak h \subseteq \mathfrak g$ considered throughout 
this section.
Let $\pr_{\mathfrak k} : \mathfrak g \to {\mathfrak k}$ be the corresponding 
projection. 
Then the condition $\alpha^K \in \mathfrak h$ (Proposition~\ref{red}) 
can be equivalently written as \hbox{$\pr_{\mathfrak k} \alpha^K = 0$}.

Denoting by $\mathcal K$ the Lie connected and simply connected matrix 
Lie group associated with the subalgebra $\mathfrak k$, we have

\begin{proposition} 
Under the above notation, the differential equations 
\begin{equation}
\label{R_comp}
\pr_{\mathfrak k} (K_x K^{-1} + K A K^{-1}) = 0, \\
\pr_{\mathfrak k} (K_y K^{-1} + K B K^{-1}) = 0
\end{equation} 
on a matrix $K \in \mathcal K$ are compatible.
\end{proposition}

\begin{pf}
Since $K \in \mathcal K$, matrices $K_x K^{-1}, K_y K^{-1}$ belong to
$\mathfrak k$ and are mapped identically under the projection 
$\pr_{\mathfrak k}$.
Hence Equations~(\ref{R_comp}) are differential equations on $K$ and, 
moreover, can be resolved with respect to $K_x,K_y$. 
Let us consider their derivatives
\begin{eqnarray*}
0 &=& D_y \pr_{\mathfrak k} (K_x K^{-1} + K A K^{-1}) 
 = \pr_{\mathfrak k} (K_{xy} K^{-1} - K B A K^{-1} + K A_y K^{-1}), 
\\
0 &=& D_x \pr_{\mathfrak k} (K_y K^{-1} + K B K^{-1}) 
 = \pr_{\mathfrak k} (K_{yx} K^{-1} - K A B K^{-1} + K B_x K^{-1}),
\end{eqnarray*}
where we have made substitutions
$\pr_{\mathfrak k} K_x K^{-1} \leadsto -\pr_{\mathfrak k}  K A K^{-1}$,
$\pr_{\mathfrak k} K_y K^{-1} \leadsto -\pr_{\mathfrak k}  K B K^{-1}$ 
according to~(\ref{R_comp}).
These equations can also be resolved with respect to $K_{xy}$ and $K_{yx}$, 
respectively.
Now one can perform the standard check that $K_{xy}$ coincides with $K_{yx}$:
$$
\pr_{\mathfrak k} (K_{xy} - K_{yx}) K^{-1}
 = \pr_{\mathfrak k} K (A_y - B_x + AB - BA) K^{-1} = 0.
$$ 
\end{pf}

\begin{definition} \rm
\label{Riccati covering}
Given a ZCR $\alpha$ of an equation $\mathcal E$ and the decomposition 
$\mathfrak g = \mathfrak h + \mathfrak k$ as above, 
we define the associated {\it generalized Riccati covering} as 
$\mathcal E \times \mathcal K \to \mathcal E$, 
assuming that the corresponding matrix of pseudopotentials
$K \in \mathcal K$ satisfies Equations~(\ref{R_comp}). 
\end{definition}

Summing up, we obtain:

\begin{corollary}
A $\mathfrak{gl}_n$-valued ZCR $\alpha$ is reducible to lower 
triangular by means of a gauge matrix from $\mathcal K$ if and only if
the generalized Riccati covering associated with the decomposition 
$\mathfrak{gl}_n = \mathfrak t_n + \mathfrak k$ has a local section.
\end{corollary}

Choosing $\mathfrak k$ to be the Lie algebra of strictly upper triangular
matrices, we have:

\begin{corollary}
A $\mathfrak{gl}_n$-valued ZCR $\alpha$ is reducible to lower triangular 
by means of a gauge matrix with nonzero principal minors if and only if 
there exists a local solution to Equations~{\rm(\ref{eqLUsln})}.
\end{corollary}

\section{Guthrie's formulation of recursion operators}

\label{GRO}

In 1994, G.A. Guthrie~\cite{G} suggested a general definition of a recursion 
operator, free of some weaknesses of the then standard definition in terms of 
integro-differential operators.
Geometrically, Guthrie's recursion operator for an equation $\mathcal E$ is 
a B\"acklund autotransformation for the linearized equation~$V\mathcal E$
(\cite{M2}). 

In geometrical terms, the linearization $V\mathcal E$ can be introduced as 
the vertical vector bundle $V\mathcal E \to \mathcal E$ with respect to the 
projection $\mathcal E \to M$ on the base manifold.

At the level of systems of PDE, the linearized system is
\begin{equation}
 \label{Veq}
F^l = 0, 
\qquad
\ell_{F^l}[U] = 0,
\end{equation}
where
\begin{equation}
 \label{lin}
\ell_{F}[U] = \sum_{k,I} \frac{\partial F}{\partial u^k_I} U^k_I
\end{equation}
(cf. the Fr\'echet derivative~\cite{Ol}), where $U^k$ are coordinates along 
the fibres of the projection $V\mathcal E \to \mathcal E$ and serve as 
additional dependent variables (`velocities' of the~$u^k$'s).
We assume summation over all $k,I$ such that the functions $F^l$ 
depend on~$u^k_I$.

Morphisms $\mathcal E \to V\mathcal E$ that are sections of the bundle 
$V\mathcal E \to \mathcal E$ are in one-to-one correspondence with local 
symmetries of the equation~$E$.
Recall that nonlocal symmetries (more precisely, their {\it shadows}
\cite{K-V}) correspond to morphisms $\tilde{\mathcal E} \to V \mathcal E$ 
over $\mathcal E$, where $\tilde{\mathcal E}$ is a covering of the original 
equation.
In full generality, Guthrie's definition includes such a covering.

Let us denote by $\tilde{V \mathcal E} \to \tilde{\mathcal E}$ the 
pullback of the vertical bundle $V \mathcal E \to \mathcal E$ along the 
covering projection $\tilde{\mathcal E} \to \mathcal E$.
Then nonlocal symmetries correspond to morphisms 
$\tilde{\mathcal E} \to \tilde{V \mathcal E}$ that are sections of the 
projection $\tilde{V \mathcal E} \to \tilde{\mathcal E}$.
In coordinates, if the covering $\tilde{\mathcal E}$ is determined by 
equations $z^j_x = f^j$, $z^j_y = g^j$, then its linearization 
$\smash{\tilde{V \mathcal E}}$ corresponds to the system 
\begin{equation}
 \label{tVeq}
F^l = 0, \quad z^j_x = f^j, \quad z^j_y = g^j, \quad \ell_{F^l}[U] = 0.
\end{equation}

\begin{definition} \label{ro} \rm (\cite{G})
A {\it recursion operator} for the system (\ref{eq}) consisting of
equations $F^l = 0$, $l = 1,\dots,s$, is given by the following data:

(1) a $\mathfrak{gl}_n$-valued zero-curvature representation 
$\bar\alpha = \bar A\,dx + \bar B\,dy$ for~$\tilde{\mathcal E}$; 

(2) two $n$-vector-valued functions $A_\circ = (A_\circ^j)$, 
$B_\circ = (B_\circ^j)$ on $\tilde{V \mathcal E}$ 
linear on the fibres (i.e., linear in the variables~$U^k_I$)
and satisfying
\begin{equation}
\label{circ}
(D_y - \bar B) A_\circ = (D_x - \bar A) B_\circ;
\end{equation}

(3) an $s \times n$-matrix-valued function $\bar C$ on~$\tilde{\mathcal E}$; 

(4) an $s$-vector-valued function $C_\circ$ on~$\tilde{V \mathcal E}$
linear on the fibres (i.e., linear in the variables~$U^k_I$). 

The following condition is supposed to hold: If $U = (U^k)$ satisfies the 
linearized equation \smash{$\tilde{V \mathcal E}$}, then so does 
$U' = L(U)$, where \smash{$L(U)^l = \bar C^l_j W^j + C_\circ^l$} and
$W^j$, $j=1,\dots,n$, are nonlocal variables of the covering
\begin{equation}
\label{K}
W^j_x = \bar A^j_i W^i + A_\circ^j, \qquad 
W^j_y = \bar B^j_i W^i + B_\circ^j,
\end{equation}
The recursion operator defined by these data will be denoted as $LK^{-1}$.
\end{definition}

Once $\bar \alpha$ is a ZCR and (\ref{circ}) holds, Equations (\ref{K}) 
determine a covering; see~\cite[Eq.~(3.2)]{G}.

Recursion operators exhibit the following form of gauge invariance:
If $S$ is a function on $E$ with values in $GL(n)$, then the data 
\begin{eqnarray}
\bar A' &=& \bar A^{S}
 = \tilde D_x S S^{-1} + S \bar A S^{-1}, \qquad
A_\circ' = S A_\circ, \nonumber \\
\bar B' &=& \bar B^{S}
 = \tilde D_y S S^{-1} + S \bar B S^{-1}, \qquad 
B_\circ' = S B_\circ, \label{geq:ro} \\
\bar C' &=& \bar C S^{-1}, \qquad C_\circ' = C_\circ \nonumber
\end{eqnarray}
(we assume matrix operations) determine the same recursion operator 
as a mapping $U \mapsto U'$.

\begin{remark} \label{gro} \rm 
One can put the definitions in a more compact form.
Let us consider $(1+n) \times (1+n)$ matrices
\begin{equation}
\label{hat}
\hat A = (\begin{array}{c|ccc}
0       & & 0      & \\
\hline
        & &        & \\
A_\circ & & \bar A & \\
        & &        & 
\end{array}),
\quad
\hat B = (\begin{array}{c|ccc}
0       & & 0      & \\
\hline
        & &        & \\
B_\circ & & \bar B & \\
        & &        & 
\end{array}).
\end{equation}
Then 
$$
\hat\alpha = \hat A\,dx + \hat B\,dy
$$ 
is a ZCR for $\tilde{V \mathcal E}$; this follows easily from 
formulas~(\ref{circ}).
Moreover, let us introduce the $s \times (1+n)$-matrix
$$
\hat C = (\begin{array}{c|ccc}
        & &        & \\
C_\circ & & \bar C & \\
        & &        & 
\end{array}).
$$
Then the above formulas (\ref{geq:ro}) of gauge invariance assume the compact 
form
\begin{eqnarray}
\hat A' &=& \hat A^{\hat S}
 = \tilde D_x \hat S \hat S^{-1} + \hat S \hat A \hat S^{-1}, \nonumber 
\\
\hat B' &=& \hat B^{\hat S}
 = \tilde D_y \hat S \hat S^{-1} + \hat S \hat B \hat S^{-1}, \label{geq:gro} 
\\
\hat C' &=& \hat C \hat S^{-1} \nonumber
\end{eqnarray}
where 
$$
\hat S = (\begin{array}{c|ccc}
1 & & 0 & \\
\hline
  & &   & \\
0 & & S & \\
  & &   & 
\end{array}).
$$

It is even possible to define a generalized recursion operator of the 
system (\ref{eq}) as consisting of a $\mathfrak{gl}_N$-valued zero-curvature 
representation $\hat\alpha = \smash{\hat A\,dx} + \smash{\hat B\,dy}$ 
for~$\smash{\tilde{V\mathcal E}}$ along with an $s \times N$-matrix-valued 
function $\smash{\hat C}$ on~$\smash{\tilde{V\mathcal E}}$,
subject to the following condition: If 
\begin{equation}
\label{KK}
\hat W^j_x = \hat A^j_i \hat W^i, \qquad 
\hat W^j_y = \hat B^j_i \hat W^i,
\end{equation}
then 
\smash{$U^{\prime l} = \hat C^l_j \hat W^j$} satisfies the linearized 
equation~$\tilde{V\mathcal E}$.

For $\hat A, \hat B$ given by formulas (\ref{hat}), the correspondence 
between $\hat W$ and $W$ is 
$$
\hat W = (\begin{array}{c}
\gamma \\
\hline
\\
\gamma W \\
\hbox{}
\end{array}),
$$
where $\gamma$ satisfies $\tilde D_x \gamma = \tilde D_y \gamma = 0$.
With $\hat S$ being an arbitrary matrix, formulas (\ref{geq:gro}) 
define a generalized gauge invariance of generalized recursion operators.
\end{remark}

Coverings~(\ref{K}) with $\bar\alpha = 0$ are associated with conservation
laws, since for them Eq.~(\ref{circ}) reads $D_y A_\circ = D_x B_\circ$.
Examples are provided by recursion operators that can be written in 
the traditional integro-differential form~(\cite{O})
$$
U^{\prime l}
 = \sum_{i = 0}^r R^{li}_k D_x^i U^k + C^l_j D_x^{-1} p^j_k U^k.
$$
Upon the obvious identification $D_x^I U^k = U^k_I$ and introduction of 
nonlocal variables $W^j = D_x^{-1} p^j_k U^k$, the Guthrie form of this 
operator is
\begin{eqnarray*}
W^j_x &=& p^{jI}_k U^k_I, \\
W^j_y &=& q^{jI}_k U^k_I, \\
U^{\prime l} &=& C^l_j W^j + R^{lI}_k U^k_I,
\end{eqnarray*}
where $p^{jI}_k U^k_I\,dx + q^{jI}_k U^k_I\,dy$ is a conservation law of 
the linearized equation $V \mathcal E$ (typically a linearized conservation 
law of the equation~$\mathcal E$;~\cite{M2}).

\begin{example} \rm \label{kdv:ro}
The Lenard recursion operator $D_{xx} + 4 u + 2 u_x D_x^{-1}$ for the KdV 
equation $u_t = u_{xxx} + 6 u u_x$ has the following Guthrie form (with
$\tilde{\mathcal E} = \mathcal E$ and 
$\smash{\tilde{V \mathcal E}} = V \mathcal E$):
\begin{eqnarray}
W_x &=& U, \nonumber \\
W_t &=& U_{xx} + 6 u U, \label{kdv:ro:a} \\
U' &=& U_{xx} + 4 u U + 2 u_x W. \nonumber
\end{eqnarray}
Indeed, if $U$ satisfies the linearized equation $V \mathcal E$, i.e., 
\begin{equation} \label{kdv:ro:b}
U_t = U_{xxx} + 6 u U_x + 6 u_x U, 
\end{equation}
then so does $U'$ (for the same $u$).

Here $W$ is a potential of the conservation law
$U\,dx + (U_{xx} + 6 u U)\,dt$ of $V\mathcal{E}$, which is a linearization 
of the conservation law $u\,dx + (u_{xx} + 3 u^2)\,dt$ of $\mathcal{E}$.
\end{example}

\subsection{Inversion of recursion operators}

A recursion operator is said to be {\it invertible} if the morphism 
$L$ of Definition~\ref{ro} is a covering.
The recursion operator $LK^{-1}$ is then simply a pair of linear coverings 
$K,L : \mathcal R \to \tilde{V \mathcal E}$, its inverse being the recursion
operator $KL^{-1}$.
Noninvertible recursion operators do exist, see Remark~\ref{ro:rem}(2).

One immediately sees that a recursion operator and its inverse are built upon 
one and the same covering $\tilde{\mathcal E}$.
In practice usually $\tilde{\mathcal E} = \mathcal E$ 
(hence the covering $\tilde{\mathcal E}$ is almost obsolete in 
Definition~\ref{ro}); however, one can simplify the ZCR~$\bar\alpha$ with
the aid of a suitably chosen covering.
Namely, given a recursion operator 
$$
\begin{array}{c@{}c@{\ }c@{\ }c@{}c}
             &         & \mathcal R \\
             & \raise.7ex \rlap{$^K$}\swarrow
             &         & \searrow\raise.7ex \llap{$^L$} \\
V \mathcal E &         &         &         & V \mathcal E
\end{array}
$$ 
associated with a ZCR $\bar\alpha$, 
the obvious pullback along a covering $p : \tilde{\mathcal E} \to \mathcal E$ 
yields a recursion operator
$$
\begin{array}{c@{}c@{\ }c@{\ }c@{}c}
             &         & p^* \mathcal R \\
             & \raise.7ex \llap{$^{p^* K}$}\swarrow
             &         & \searrow\raise.7ex \rlap{$^{p^* L}$} \\
\tilde{V \mathcal E} = 
p^*V\mathcal E
             &         &         &         & p^*V\mathcal E
                                             = \tilde{V \mathcal E},
\end{array}
$$
which is associated with the pullback $p^*\bar\alpha$.

For instance, let $\tilde{\mathcal E} \to \mathcal E$ be the 
trivializing covering for $\bar\alpha$.
Then, after suitable transformation (\ref{geq:ro}), we have
$p^*\bar\alpha = 0$,
whence the recursion operator becomes integro-differential of first order in 
$D^{-1}$.
Hence a possible way of conversion of recursion operators from Guthrie's 
to Olver's form, mentioned in the Introduction.
This approach was used in the work by Guthrie 
and Hickman~\cite{G-H} who, by using formal power series in the spectral 
parameter $\lambda$, were able to describe large algebras of nonlocal 
symmetries of the KdV equation resulting from iterated application of the 
inverse recursion operator.

Alternatively, $\tilde{\mathcal E} \to \mathcal E$ can be a covering such 
that $p^* \bar\alpha$ is strictly lower triangular (belongs to 
$\mathfrak t^{(1)}$).
Then the covering (\ref{K}) will be abelian by a similar argument as in 
Proposition~\ref{p:triv_abel} and the recursion operator 
will be integro-differential of order $\le s$ in $D^{-1}$.

Let us now turn back to recursion operators $L K^{-1}$ with a general covering 
$\bar\alpha$.
One usually observes that for systems 
$\mathcal E$ integrable in the sense of soliton theory the covering $K$ 
is of a very special form, which is described in the following proposition:

\begin{proposition}
Let $\alpha = A\,dx + B\,dy$ be a $\mathfrak g$-valued ZCR of equation 
$\mathcal E$.
Then the trivial vector bundle 
$\mathfrak g \times V \mathcal E \to V \mathcal E$ carries a covering
structure determined by the condition that an arbitrary element $W$ 
of the Lie algebra $\mathfrak g$ be subject to equations
\begin{equation} 
\label{W}
W_x = [A,W] + \ell_A[U], \qquad
W_y = [B,W] + \ell_B[U].
\end{equation}
\end{proposition}

Otherwise said, the associated ZCR $\bar\alpha$ coincides with the 
adjoint representation of the ZCR $\alpha$, while $A_\circ = \ell_A[U]$, 
$B_\circ = \ell_B[U]$.

\begin{pf}
The validity of formulas (\ref{circ}) follows from the fact that 
$A \mapsto \ell_A[U]$ is a differentiation.
\end{pf}

Taking account of the last proposition, we arrive at the following 
construction, which converts a recursion operator from Guthrie's form to 
Olver's form provided the covering $K$ is of the type~(\ref{W}).

\begin{construction} \rm
\label{Z}
Step 1. \
Construct the generalized Riccati covering 
(Definition~\ref{Riccati covering})
$\mathcal E'$ over $\mathcal E$ such that $\alpha' := \alpha^{H}$ is lower 
triangular, where $H$ is the matrix~(\ref{H}).

Step 2. \
Let $a_{ii}'$, $b_{ii}'$ be the diagonal entries of the lower triangular 
matrices $A^{H}$, $B^{H}$, respectively. 
Then $a_{ii}'\,dx + b_{ii}'\,dy$ are conservation laws; 
if they are nontrivial, then construct the abelian covering $\mathcal E''$ 
over $\mathcal E'$ with the corresponding potentials~$z_i$.

Step 3. \
Compute $S = ZH$, where $Z$ is the diagonal matrix 
diag$({\mathrm e}^{-z_i})$.
Obviously, $\alpha'' := \alpha^S$ is then strictly lower triangular,
and so is its adjoint representation 
$$
\overline{\alpha''} = \bar\alpha^{\bar S}, 
$$
where $\bar S$ is the image of $S$ in the adjoint representation of the 
group~$G$.
The $x$-part of the resulting recursion operator given by 
formulas~(\ref{geq:ro}) will be expressible in terms of inverse total 
derivatives~$D_x^{-1}$.

Step 4 (optional). \
Let us consider the compact form~(\ref{hat}) 
of the recursion operator, which now takes values in the algebra
\smash{$\mathfrak{t}_{n+1}^{(1)}$} of strictly lower triangular matrices of
dimension $n+1$.
Choosing appropriately a lower triangular gauge matrix $\hat S$ with units 
on the diagonal, one can, in principle, further simplify the formulas.
\end{construction}

If omitting Step 2, the recursion operator will be expressible in terms of
inverses $(D_x - a'_{ii})^{-1}$.

\begin{remark} \rm \label{ro:rem}
(1)
Let $R$ be a conventional recursion operator of an integrable system, let id 
denote the identity map.
As a rule, the inverse recursion operator $(R + \lambda\,\hbox{id})^{-1}$ 
in the Guthrie form includes a $\lambda$-dependent ZCR $\bar\alpha$.
The parameter $\lambda$ can be usually identified with the spectral parameter 
of the standard ZCR of the system.

(2)
Let us recall that the formulas (\ref{W}) can serve as a starting point 
of a method to find the inverse recursion operator of an integrable system 
without finding the conventional recursion operator first.
One simply computes all morphisms $\mathcal R \to V \mathcal E$, where 
$\mathcal R$ is the covering determined by~(\ref{W}).
Recently the procedure has been applied to the stationary 
Nizhnik--Veselov--Novikov equation, see~\cite{M-S}.
Remarkably enough, the so obtained recursion operator turned out 
to be noninvertible for the zero value of the spectral parameter~$\lambda$.
Two examples of such computation can be found below.
\end{remark}

\section{Examples}

\begin{example} \rm \label{kdv:ro:inv}
Continuing Example \ref{kdv:ro}, let us invert the Lenard operator.
The result is, of course, well known (Guthrie and Hickman~\cite{G-H}, 
Lou~\cite{L3,L-C}).

Instead of tedious inversion of the operator given by formulas
(\ref{kdv:ro:a}) and (\ref{kdv:ro:b}), we compute it from scratch.
We start with the standard $\mathfrak{sl}_2$-valued ZCR
$$
\alpha
 = (\begin{array}{rc}
0 & u \\
-1 & 0 
\end{array}) dx
 + (\begin{array}{rc}
u_x & u_{xx} + 2 u \\
-2 u & -u_x
\end{array}) dy
$$  
of the KdV equation with the spectral parameter set to zero. 
Using (\ref{K}) and~(\ref{W}) with $\mathfrak{sl}_2$ parametrized as
$$
(\begin{array}{cr}
Q & P \\
R & -Q
\end{array}),
$$ 
we get the following formulas for the covering $K$:
\begin{eqnarray*}
(\begin{array}{c}
P \\ Q \\ R 
\end{array})_{\!x}
&=&
(\begin {array}{ccc} 
0 & -2 u & 0 \\ 
1 & 0 & u \\ 0 & \llap{$-$}2 & 0
\end{array}) \!\!
(\begin{array}{l}
P \\ Q \\ R 
\end{array})
 + 
(\begin{array}{c}
U \\
0 \\
0
\end{array})\!,
\\
(\begin{array}{c}
P \\ Q \\ R 
\end{array})_{\!t}
&=&
(\begin {array}{ccc} 
2 u_x & -2 u_{xx} - 4 u^2 & 0 \\ 
2 u & 0 & u_{xx} + 2 u^2 \\ 
0 & -4 u & -2 u_x
\end{array}) \!\!
(\begin{array}{l}
P \\ Q \\ R 
\end{array})
 + 
(\begin{array}{c}
U_{xx} + 4 u U \\
U_x \\
\llap{$-$}2 U
\end{array})\!.
\end{eqnarray*}
Here $U$ denotes a symmetry of the KdV equation, i.e., satisfies 
the linearized KdV equation (\ref{kdv:ro:b}).
Then one easily finds that $U' = Q$ satisfies the same linearized KdV 
equation (\ref{kdv:ro:b}) as well, i.e.,  yields a recursion operator for 
the KdV equation.
It is a matter of routine to check that this operator is the inverse of the
Lenard operator.
Moreover, it follows that $K : \mathcal R \to V \mathcal E$, originally 
given by $U = U'_{xx} + 4 u U' + 2 u_x W$, constitutes a
three-dimensional covering (with nonlocal variables $U,U_x$ and~$W$).

According to Construction~\ref{Z}, to express the inverse recursion operator 
in terms of $D_x^{-1}$, all we need is to make the ZCR $\bar\alpha$ 
strictly lower triangular.
As the first step we build up a covering $\mathcal E' \to \mathcal E$ with 
the quadratic pseudopotential $h = h_{11}$ defined by Eq.~(\ref{qupsp}), i.e., 
\begin{eqnarray*}
h_x &=& -h^2 - u, \\
h_t &=& -2 u h^2 + 2 u_x h - u_{xx} - 2 u^2.
\end{eqnarray*}
Then, using the gauge matrix
$$
H = (\begin {array}{cc} 1 & h \\ 0 & 1\end{array}),
$$
we get the lower triangular ZCR
$$
\alpha' = \alpha^H = 
(\begin {array}{cc}
 -h & 0 \\
 -1 & h
\end{array}) dx +
(\begin {array}{cc}
 u_x - 2 u h & 0 \\
 -2 u & -u_x + 2 u h
\end{array}) dy
$$
with $-h, h$ on the diagonal.
As the second step, we construct the abelian covering 
$\mathcal E'' \to \mathcal E'$ with the potential $z$ satisfying
$$
z_x = -h, \qquad
z_y = u_x - 2 u h.
$$
The gauge matrix
$$
Z = (\begin{array}{cc} 
{\mathrm e}^{-z} & 0 \\ 
0 & {\mathrm e}^z
\end{array})
$$
then gives the strictly lower triangular ZCR
$$
\alpha'' = \alpha^{ZH}
 = (\begin {array}{cc}
 0 & 0 \\
 -\mathrm e^{2z} & 0
\end{array}) dx +
(\begin {array}{cc}
 0 & 0 \\
 -2 \mathrm e^{2z} u & 0
\end{array}) dy.
$$

In the third step, we combine the above gauge matrices into one
and compute its adjoint representation:
$$
S = (\begin {array}{cc} 
\mathrm e^{-z} & h \mathrm e^{-z} \\ 0 & \mathrm e^{z}
\end{array}),
\qquad
\bar S = (\begin {array}{ccc} 
 \mathrm e^{-2 z} & -2 h \mathrm e^{-2 z} & -h^2 e^{-2 z} \\
 0 & 1 & h \\
 0 & 0 & \mathrm e^{2 z}
\end{array}).
$$
Acting by $\bar S$ on our operator, we get
\begin{eqnarray*}
(\begin{array}{c}
P \\ Q \\ R 
\end{array})_{\!x}
&=&
(\begin {array}{ccc} 
0 & 0 & 0 \\ e^{2 z} & 0 & 0 \\ 0 & -2 e^{2 z} & 0
\end{array}) \!\!
(\begin{array}{l}
P \\ Q \\ R 
\end{array})
 + 
(\begin{array}{c}
\mathrm e^{-2 z} 
U' \\ 0 \\ 0
\end{array})\!,
\\
(\begin{array}{c}
P \\ Q \\ R 
\end{array})_{\!t}
&=&
(\begin {array}{ccc} 
0 & 0 & 0 \\ 2 u e^{2 z} & 0 & 0 \\ 0 & \llap{$-$}4 u e^{2 z} & 0
\end{array}) \!\!
(\begin{array}{l}
P \\ Q \\ R 
\end{array})
\\
 && + 
(\begin{array}{c}
\mathrm e^{-2 z} U'_{xx} - 2 \mathrm e^{-2 z} h U'_x
 + (2 h^2 + 4 u) \mathrm e^{-2 z} U' \\
U'_x  - 2 h U' \\
\llap{$-$}2 \mathrm e^{2 z} U'
\end{array}), \\
U &=& Q - h \mathrm e^{-2 z} R.
\end{eqnarray*}
Rewritting the $x$-part in terms of inverse total derivatives $D^{-1}$, 
we get 
$P = D^{-1} (\mathrm e^{-2 z} U)$,
$Q = D^{-1} (\mathrm e^{2 z} P)$, 
$R = -2 D^{-1} (\mathrm e^{-2 z} Q)$, 
hence
$$
U 
 = D^{-1} \mathrm e^{2 z} D^{-1} \mathrm e^{-2 z} U
 - h \mathrm e^{-2 z} D^{-1} \mathrm e^{-2 z} D^{-1} \mathrm e^{2 z}
    D^{-1} \mathrm e^{-2 z} U.
$$
This is the well-known result~\cite{G-H,L3,L-C}, since
$U' = Q - h \mathrm e^{-2 z} R = -\frac 12 D_x(R/h^2)$ and
$z_{xx} = z_x^2 + u$.

The optional fourth step does not bring any significant improvement.
\end{example}

\begin{example} \rm \label{tzitzeica:ro:inv}
Let us consider the Tzitz\'eica equation~\cite{Tz}
$$
u_{xy} = \mathrm e^{u} - \mathrm e^{-2 u},
$$
later rediscovered as a member of the Zhiber--Shabat classification~\cite{Zh-S}.
Its ZCR
\begin{equation}
\label{tz:zcr}
\alpha = 
(\begin{array}{@{\quad}ccc} 
\llap{$-$}u_x & 0 & \lambda \\ 
\lambda & u_x & 0 \\ 
0 & \lambda & 0 
\end{array}) dx
 + 
(\begin{array}{ccc} 
0 & {\mathrm e^{-2 u}}/{\lambda} & 0 \\ 
0 & 0 & {\mathrm e^{u}}/{\lambda} \\ 
{\mathrm e^{u}}/{\lambda} & 0 & 0 
\end{array}) dy
\end{equation}
as well as the B\"acklund transformation were essentially found by 
Tzi\-tz\'eica himself.

One could invert the known recursion operator~\cite{S-Sh}, but it is easier 
to compute the inverse recursion operator directly by the procedure outlined 
in Remark~\ref{ro:rem}(2).
Namely, we consider the eight-dimensional covering (\ref{W}),  
where \smash{$\bar A, \bar B, A_\circ$} and $B_\circ$ are found from the 
formula~(\ref{tz:zcr}) to be
\begin{eqnarray*}
\bar A &=& 
(\begin{array}{@{\quad}ccc@{\quad}c@{\quad}c@{\quad\,}c@{\quad}cc}
 0 & \llap{$-$}\lambda & 0 & 0 & 0 & 0 & \lambda & 0 \\
 0 & -2 u_x & \llap{$-$}\lambda & 0 & 0 & 0 & 0 & \lambda \\ 
 \llap{$-$}2 \lambda & 0 & -u_x & 0 & \llap{$-$}\lambda & 0 & 0 & 0 \\
 \lambda & 0 & 0 & 2 u_x & \llap{$-$}\lambda & 0 & 0 & 0 \\
 0 & \lambda & 0 & 0 & 0 & \llap{$-$}\lambda & 0 & 0 \\
 0 & 0 & \lambda & \llap{$-$}\lambda & 0 & u_x & 0 & 0 \\
 0 & 0 & 0 & \lambda & 0 & 0 & u_x & \llap{$-$}\lambda \\
 \lambda & 0 & 0 & 0 & 2 \lambda & 0 & 0 & -u_x 
\end{array}), 
\\
\bar B &=& (\begin{array}
 {@{\quad\!}cccccc@{\hskip 1.5ex}c@{\hskip 1.5ex}c}
 0 & 0 & \llap{$-$}{\mathrm e^{u}}/{\lambda} & {\mathrm e^{-2 u}}/{\lambda} & 
  0 & 0 & 0 & 0 \\
 \llap{$-$}{\mathrm e^{-2 u}}{\lambda} & 0 & 0 & 0 &
  {\mathrm e^{-2 u}}/{\lambda} & 0 & 0 & 0 \\
 0 & \llap{$-$}{\mathrm e^{u}}/{\lambda} & 0 & 0 & 0 &
  {\mathrm e^{-2 u}}/{\lambda} & 0 & 0 \\
 0 & 0 & 0 & 0 & 0 & \llap{$-$}{\mathrm e^{u}}/{\lambda} &
  {\mathrm e^{u}}/{\lambda} & 0 \\ 
 0 & 0 & 0 & \llap{$-$}{\mathrm e^{-2 u}}/{\lambda} & 0 & 0 & 0 &
  {\mathrm e^{u}}/{\lambda} \\
 \llap{$-$}{\mathrm e^{u}}/{\lambda} & 0 & 0 & 0 &
  \llap{$-$}2 {\mathrm e^{u}}/{\lambda} & 0 & 0 & 0 \\ 
 2 {\mathrm e^{u}}/{\lambda} & 0 & 0 & 0 & {\mathrm e^{u}}/{\lambda} & 0 & 0 &
  0 \\
 0 & {\mathrm e^{u}}/{\lambda} & 0 & 0 & 0 & 0 &
  \llap{$-$}{\mathrm e^{-2 u}}/{\lambda} & 0 
\end{array})\!,
\\
A_\circ &=& 
(\begin{array}{c}
 -U_x \\ 0 \\ 0 \\ 0 \\ U_x \\ 0 \\ 0 \\ 0
\end{array}),
\qquad
B_\circ = (\begin{array}{c}
 0 \\ -2 {\mathrm e^{-2 u} U}/{\lambda} \\ 0 \\ 0 \\ 0 \\ 
 {\mathrm e^{u} U}/{\lambda} \\ {\mathrm e^{u} U}/{\lambda} \\ 0
\end{array}),
\end{eqnarray*}
$W$ being a column
$(W_{11},W_{12},W_{13}, W_{21},W_{22},W_{23}, W_{31},W_{32})^\top$
of pseudopotentials.
One easily finds that $W_{11} - W_{22}$ is a symmetry of the Tzitz\'eica 
equation if so is~$U$.
We have obtained the `inverse' recursion operator of the 
Tzitz\'eica equation in the Guthrie form.

Let us express it in terms of $D_x^{-1}$.
As the first step we introduce pseudopotentials
$p,q,r$ satisfying
$$
\begin{array}{l@{\qquad}l}
p_x = \lambda  p^2 - 2 p u_x - \lambda  q, & \displaystyle
p_y = \frac{\mathrm e^{u}}{\lambda} p q
        - \frac{1}{\mathrm e^{2u} \lambda}, \\\noalign{\vskip 1ex}
q_x = \lambda  p q - q u_x - \lambda, & \displaystyle
q_y = \frac{\mathrm e^{u}}{\lambda} (q^2 - p), \\\noalign{\vskip 1ex}
r_x = - \lambda  p r + \lambda  q + \lambda  r^2 + u_x r, & \displaystyle
r_y = \frac{\mathrm e^{u}}{\lambda} (-p r^2 + q r - 1).
\end{array}
$$
to make the ZCR~(\ref{tz:zcr}) lower triangular by providing a solution to 
Equations~(\ref{ULgln}).
Indeed, acting on $\alpha$ by the gauge matrix 
$$
H = 
(\begin{array}{ccc}
 1 & p & q \\ 0 & 1 & r \\ 0 & 0 & 1 
\end{array})
$$
we get
\begin{eqnarray*}
\alpha^H &=& 
(\begin{array}{ccc} -u_x + \lambda  p & 0 & 0 \\ \lambda & 
u_x - \lambda  p + \lambda  r & 0 \\ 0 & \lambda & -\lambda  r 
\end{array})dx \\
 &&+ 
(\begin{array}{ccc} 
 {\mathrm e^{u} q}/{\lambda} & 0 & 0 \\
 {\mathrm e^{u} r}/{\lambda} & -{\mathrm e^{u} pr}/{\lambda} & 0 \\ 
 {\mathrm e^{u}}/{\lambda} & -{\mathrm e^{u} p}/{\lambda} & 
  {\mathrm e^{u}} (p r - q)/{\lambda} 
\end{array})dy.
\end{eqnarray*}
In the second step we remove the diagonal.
To this end we introduce pseudopotentials $s,t$ by
$$
\begin{array}{l@{\qquad}l}
s_x = -u_x + \lambda  p, & \displaystyle
s_y = \frac{\mathrm e^{u}}{\lambda} q, \\\noalign{\vskip 1ex plus 1ex}
t_x = u_x - \lambda  p + \lambda  r, & \displaystyle
t_y = -\frac{\mathrm e^{u}}{\lambda} p r.
\end{array}
$$
Acting on $\alpha^H$ by the gauge matrix 
$$
Z = (\begin{array}{ccc}
 \mathrm e^{-s} & 0 & 0 \\
 0 & \mathrm e^{-t} & 0 \\
 0 & 0 & \mathrm e^{s + t} 
\end{array})
$$
we finally get
\begin{eqnarray*}
\alpha^{ZH} &=& 
(\begin{array}{ccc} 0 & 0 & 0 \\ 
\lambda  \mathrm e^{s - t} & 0 & 0 \\ 0 & 
\lambda  \mathrm e^{s + 2 t} & 0 \end{array}) dx \\
&& +
(\begin{array}{ccc}
 0 & 0 & 0 \\ 
 {\mathrm e^{u + s - t} r}/{\lambda} & 0 & 0 \\ 
 {\mathrm e^{u + 2 s + t}}/{\lambda} &
    -{\mathrm e^{u + s + 2 t} p}/{\lambda} & 0 
\end{array}) dy.
\end{eqnarray*}
Denoting $S = ZH$, we compute the adjoint representation $\bar S$ to be
$$
\begin{array}{l}
\bar S = (\begin{array}{ccccc}
 \mathrm e^{-2 s - t} & -\mathrm e^{-2 s - t} r & \mathrm e^{-2 s - t} p & 
  \mathrm e^{-2 s - t}(p r - 2 q) & -\mathrm e^{-2 s - t}(p r + q) \\ 
 0 & \mathrm e^{-s + t} & 0 & -\mathrm e^{-s + t} p & 
  \mathrm e^{-s + t} p \\
 0 & 0 & \mathrm e^{-s - 2 t} & -\mathrm e^{-s - 2 t} r &
  \llap{$-$}2 \mathrm e^{-s - 2 t} r \\
 0 & 0 & 0 & 1 & 0 \\
 0 & 0 & 0 & 0 & 1 \\
 0 & 0 & 0 & 0 & 0 \\
 0 & 0 & 0 & 0 & 0 \\
 0 & 0 & 0 & 0 & 0 
\end{array} \right.
\\
\qquad\qquad\qquad\qquad\quad \left.
\begin{array}{ccc}
  \mathrm e^{-2 s - t} p (p r - q) & -\mathrm e^{-2 s - t}q r  &
   \mathrm e^{-2 s - t} q (p r - q) \\ 
  \llap{$-$}\mathrm e^{-s + t} p^2 & \mathrm e^{-s + t} q &
   \llap{$-$}\mathrm e^{-s + t} p q \\
  \mathrm e^{-s - 2 t}(p r - q) & -r^2 \mathrm e^{-s - 2 t} &
   \mathrm e^{-s - 2 t}(p r - q) r \\
  p & 0 & q \\
  \llap{$-$}p & r & \llap{$-$}p r \\
  \mathrm e^{s - t} & 0 & \mathrm e^{s - t} r \\
  0 & \mathrm e^{s + 2 t} & \llap{$-$}\mathrm e^{s + 2 t} p \\
  0 & 0 & \mathrm e^{2 s + t} 
\end{array}).
\end{array}
$$
Acting by $\bar S$ on the above recursion operator we get
$$
\bar A^{\bar S} = 
(\begin{array}{@{\quad}cc@{\quad\ }cc@{\ \ }cc@{\quad\ }cc}
 0 & 0 & 0 & 0 & 0 & 0 & 0 & 0 \\ 
 \llap{$-$}\lambda \mathrm e^{s + 2 t} & 0 & 0 & 0 & 0 & 0 & 0 & 0 \\ 
 \lambda  \mathrm e^{s - t} & 0 & 0 & 0 & 0 & 0 & 0 & 0 \\
 0 & \llap{$-$}\lambda \mathrm e^{s - t} & 0 & 0 & 0 & 0 & 0 & 0 \\
 0 & \lambda  \mathrm e^{s - t} & \llap{$-$}\lambda \mathrm e^{s + 2 t} & 
   0 & 0 & 0 & 0 & 0 \\
 0 & 0 & 0 & \lambda \mathrm e^{s - t} & \llap{$-$}\lambda \mathrm e^{s - t} &
   0 & 0 & 0 \\
 0 & 0 & 0 & \lambda \mathrm e^{s + 2 t} & 2 \lambda \mathrm e^{s + 2 t} &
   0 & 0 & 0 \\
 0 & 0 & 0 & 0 & 0 & \lambda \mathrm e^{s + 2 t} &
  \llap{$-$}\lambda \mathrm e^{s - t} & 0 
\end{array})
$$
and
$$
\bar S A_\circ = 
(\begin{array}{c}
 \mathrm e^{-2 s - t}(-2 p r + q) U_x \\ 
 2 \mathrm e^{-s + t} p U_x \\
 \llap{$-$} \mathrm e^{-s - 2 t} r U_x \\
 \llap{$-$}U_x \\ 
 U_x \\
 0 \\
 0 \\
 0 
\end{array})
$$
(we omit the matrices $\bar B^{\bar S}$ and $B_\circ$).

Thus, the inverse recursion operator for the Tzitz\'eica equation in
terms of $D^{-1}$ is
$$
V =
 W_{21} - W_{22}
 - 2 \mathrm e^{-s + t} p W_{23}
 + \mathrm e^{-s - 2 t} r W_{31}
 + \mathrm e^{-2 s - t} (2 p r - q) W_{32},
$$
where
\begin{eqnarray*}
W_{11} &=& D^{-1}[\mathrm e^{-2 s - t} (-2 p r + q) U_x],
\\
W_{12} &=& D^{-1}
[2 \mathrm e^{-s + t} p U_x - \mathrm e^{s + 2 t} \lambda  W_{11}],
\\
W_{13} &=& D^{-1}
[-\mathrm e^{-s - 2 t} r U_x + \mathrm e^{s - t} \lambda  W_{11}],
\\
W_{22} &=& D^{-1}
[U_x + \mathrm e^{s - t} \lambda W_{12} - \mathrm e^{s + 2 t} \lambda W_{13}],
\\
W_{21} &=& D^{-1}
[-U_x - \mathrm e^{s - t} \lambda W_{12}],
\\
W_{31} &=& D^{-1}
[\mathrm e^{s + 2 t} \lambda (W_{21} + 2 W_{22})],
\\
W_{23} &=& D^{-1}
[-\mathrm e^{s - t} \lambda (-W_{21} + W_{22})],
\\
W_{32} &=& D^{-1}
[\lambda  (\mathrm e^{s + 2 t} W_{23} - \mathrm e^{s - t} W_{31})].
\end{eqnarray*}
\end{example}

\section*{Acknowledgements}

This work was very much influenced by discussions with A.~Sergyeyev 
and V.V.~Sokolov.
The support from the grant MSM:J10/98:192400002 is gratefully acknowledged.

\end{document}